\documentclass{elsart}

\usepackage{graphicx}
\usepackage{epsfig,rotating}
\usepackage[dvips]{color}

\def\be{\begin{equation}}
\def\ee{\end{equation}}
\def\ba{\begin{eqnarray}}
\def\ea{\end{eqnarray}}

\def\b{\beta}
\def\e{\epsilon}
\def\o{\omega}
\def\z{\zeta}
\def\G{\Gamma}

\def\lsim{\raise0.3ex\hbox{$\;<$\kern-0.75em\raise-1.1ex\hbox{$\sim\;$}}}
\def\gsim{\raise0.3ex\hbox{$\;>$\kern-0.75em\raise-1.1ex\hbox{$\sim\;$}}}

\begin{document}

\begin{frontmatter}
\title{Constraining cosmic superstrings\\ with dilaton emission}

\author[MPI,INR]{E.~Babichev}
\author[MPI]{and M.~Kachelrie\ss}

\address[MPI]{Max-Planck-Institut f\"ur Physik
(Werner-Heisenberg-Institut),
F\"ohringer Ring 6, 80805 M\"unchen, Germany}
\address[INR]{INR RAS, 60th October Anniversary prospect 7a,
  117312 Moscow, Russia}

\date{\today}

\begin{abstract}
Brane inflation predicts the production of cosmic
superstrings with tension $10^{-12}\lsim G\mu\lsim 10^{-7}$. 
Superstring theory predicts also the existence of a dilaton with a mass
that is at most of the order of the gravitino mass. We show that the
emission of dilatons imposes severe constraints on the allowed
evolution of a cosmic superstring network. In particular, the
detection of gravitational wave burst from cosmic superstrings by LIGO
is only possible if the typical length of string loops is much smaller
than usually assumed. 

\begin{small}
PACS: 
04.30.Db,        
11.27.+d,        
98.80.Cq         
\hfill Preprint MPP-2005-11
\end{small}
\end{abstract}
\end{frontmatter}

\section{Introduction}

The inflationary paradigm has become one of the corner stones of modern
cosmology. While this paradigm solves many puzzles of the ``old'' big
bang theory, no convincing theoretical framework exists at present for
the inflaton field and its potential. In the brane
inflation scenario~\cite{dt98}, the inflaton is identified as a mode
controlling the separation of two branes. Inflation ends when the branes
collide and heat-up the Universe. An intriguing prediction of most
scenarios of brane inflation is the copiously production of cosmic
superstrings~\cite{ss}.

One of the few generic predictions of superstring theory is the
existence of the dilaton supermultiplet in its spectrum. Independent of
the particular mechanism of supersymmetry breaking, any realistic
string theory model should led to a low-energy theory with softly
broken supersymmetry at the TeV scale in order to solve the hierarchy
problem. This ensures together with other phenomenological constraints
that the dilaton mass is at most of the order of the gravitino mass
$m_{3/2}$~\cite{BdC}. In hidden sector models of supersymmetry
breaking $m_{3/2}$ is around the electroweak scale, while in
gauge-mediated supersymmetry breaking models $m_{3/2}$ could be as
low as $0.01$~keV.

Cosmological consequences of dilatons in string theories have been
already widely discussed~\cite{BdC,Asaka,DamVil97}. The purpose of
this work is the investigation of dilaton emission from a network of cosmic
superstrings. We derive stringent limits on the tension of the strings
as function of the dilation mass and discuss the maximal
gravitational wave signal from cosmic strings compatible with these
limits.

\section{Dilaton radiation from a cosmic superstring network}
In this section, we reanalyze the emission of dilatons by a 
cosmic string network. Our main aim is the generalization of the
analysis of Damour and Vilenkin in Ref.~\cite{DamVil97} to the case of
a more general network evolution, where the energy losses of
strings as well as their reconnection probability can deviate from
their ``standard'' values.

Let us first recall the main quantities describing the standard
evolution of a string network.  The network is characterized 
by the typical length scale $L^{\rm st}(t)\sim t/\sqrt{\z}$,
where $\z\sim 1/4$ is a parameter that has to be determined in
numerical simulations. Another characteristic quantity of the
cosmic string network is the parameter $\b$ characterizing the 
length $l$ of loops which are chopped-off long strings,
\be
\label{l}
l(t)\sim \b t \,.
\ee
In the standard scenario, this coefficient $\b$ is determined
by the gravitational radiation losses from loops,
\be
\label{bst} 
\b^{\rm st}\sim \Gamma G\mu \,.
\ee
Here, $\Gamma$ is a numerical coefficient of the order 50. 
At any given moment of time $t$, the typical size of loops is given by
inserting into Eq.~(\ref{l}) the appropriate value of $\b$. Then the density
of such loops $n^{\rm st}$  can be written as 
\be
\label{nst}
n^{\rm st}\sim \frac{\z}{\G G\mu\,t^3} \,.
\ee

The evolution of a string network can be modified in two ways:
First, the probability of intercommuting of intersecting strings
may be different from one, $p\le 1$, as it happens for cosmic
superstrings~\cite{Jackson:2004zg}.  
Second, the value of the coefficient $\beta$ can differ from the one
usually assumed, $\beta\neq \G G\mu$. Note that the first modification
of the properties of cosmic strings ($p\neq 1$) 
is due to the difference between superstrings and topological strings, 
while the possible deviation of $\beta$ from the standard value 
applies for both type of strings. Introducing a non-standard value for
$\beta$ is inspired by Ref.~\cite{SO} where it was argued that $\beta$
could be much smaller than it was originally assumed. It is convenient to 
introduce the ratio
\be
\label{eps}
\e=\frac{\b}{\b^{\rm st}}.
\ee
Following \cite{DamVil04}, one can find the resulting  
differences in the properties of strings for the two extreme cases
$\e\ll 1$ and $\e\gg 1$. In the latter case, the typical length of
loops is given by 
\be
\label{l1}
l_1\sim \G G\mu\,t,
\ee
and their density is
\be
\label{n1}
n_1\sim\frac{\e^{1/2}\z}{p\,\G G\mu\, t^3} \,.
\ee
(Note that we are interested in the radiation domination epoch, while 
in \cite{DamVil04} the matter domination epoch was considered. This
leads to different expressions for the loop density.) In the opposite
case, $\e\ll 1$, one obtains as typical length of loops
\be
\label{l2}
l_2\sim \beta\, t \,,
\ee
and as density of loops
\be
\label{n2}
n_2\sim\frac{\z}{p\,\G G\mu\, t^3} \,.
\ee

Let us now turn to the calculation of the number of dilatons radiated
from a cosmic superstring network. In Ref.~\cite{DamVil97}, it was shown 
that a single loop radiates dilatons of frequency $\o$ with the rate 
\be
\label{dotN}
\dot{N_\phi}=\G_\phi\, \alpha^2\, G\mu^2/\o.
\ee
The coefficient $\G_\phi$ does not depend on the loop size and is
typically $\G_\phi\sim 13$, while
$\alpha=\partial\ln\sqrt{\mu(\phi)}/\partial\phi$ measures the
strength of the coupling of the dilaton to the string relative to the
gravitational strength. To simplify the analysis, we assume that
all dilatons are emitted at the same fundamental frequency 
$\o=4\pi/l$. Then the total number of radiated dilatons can be 
written as $N_\phi\sim\dot N_\phi \tau$, where $\tau$ is the 
decay time of a loop. For $\e\gg 1$, this time is simply 
$\tau\sim t$. Since the typical length of loops is for $\e\ll 1$
smaller by a factor $\e$, the decay time of a typical loop is by the
same factor suppressed, $\tau\sim \e t$. As result we obtain from
Eq.~(\ref{dotN}) as total number of emitted dilatons from a single loop
\be
  \label{N}
    N\sim \left\{\begin{array}{lcl}
     (4\pi)^{-1}\G \G_\phi \alpha^2\, G^2\,\mu^3 t^2, & & \e\gg 1, \\
     (4\pi)^{-1}\G \G_\phi \alpha^2\, G^2\,\mu^3 t^2 \e^2, & & \e\ll 1.\\
    \end{array} \right. \,
\ee
It is convenient to introduce further the relative abundance of
dilatons, $Y_\phi=n_\phi(t)/s(t)$, where $n_\phi(t)$ is the density of
dilatons and $s(t)$ the entropy density. In the radiation epoch, the
entropy density is given by 
\be
\label{s}
s(t)=0.0725 [\mathcal{N}(t)]^{1/4} \left(\frac{M_{\rm Pl}}{t}\right)^{3/2},
\ee
where $\mathcal{N}(t)$ is the effective number of spin degrees of 
freedom and $M_{\rm Pl}$ the Planck mass. Loops which decay at time
$t$ contribute to $Y_\phi$ the following abundance of dilatons,
\be
  \label{Y}
    Y_\phi(t)\sim\frac{n\,N}{s}\sim \left\{\begin{array}{lcl}
     p^{-1}\,\e^{1/2}\,\z \,\G_\phi \alpha^2\, (G\mu)^2\,(M_{\rm Pl} t)^{1/2} \mathcal{N}^{-1/4}, & & \e\gg 1, \\
     p^{-1}\,\e^{2}\,\z \,\G_\phi \alpha^2\, (G\mu)^2\,(M_{\rm Pl} t)^{1/2} \mathcal{N}^{-1/4}, & & \e\ll 1.\\
    \end{array} \right. \,
\ee
The expression (\ref{Y}) differ from the one obtained in Ref.~\cite{DamVil97}
by a factor $p^{-1}$ for $\e\gg 1$ and by a factor  $p^{-1}\,\e^{3/2}$ 
for $\e\ll 1$. Equation~(\ref{Y}) is valid as long as the loop sizes are
smaller than the critical size $l_c=4\pi/m_\phi$, or $t_c\lsim
4\pi/(\G\,G\mu\,m_\phi)$ for $\e\gg 1$ and  $t_c\sim
4\pi/(\e\,\G\,G\mu\,m_\phi)$ for $\e\ll 1$. 
Substituting these times into  Eq.~(\ref{Y}) we find
\be
  \label{Y1}
    Y_\phi(t)\sim \left\{\begin{array}{lcl}
     20\, p^{-1}\,\e^{1/2}\,\alpha^2\, (G\mu)^{3/2}\,(M_{\rm Pl} /m_\phi)^{1/2}, & & \e> 1, \\
     20\, p^{-1}\,\e^{3/2}\,\alpha^2\, (G\mu)^{3/2}\,(M_{\rm Pl} /m_\phi)^{1/2} & & \e< 1.\\
    \end{array} \right. \,
\ee
In the last expression we replaced also the strong inequalities 
$\e\gg 1$ and $\e\ll 1$ by $\e>1$ and $\e<1$ to include 
the range $\e\sim 1$.

The derivation of the expression (\ref{Y1}) has been made 
for the scaling regime of the network. This assumes in particular that 
plasma friction is negligible. Since for times 
$t_*\lsim 1/M_{\rm Pl}\,(G\mu)^2$ the motion of cosmic strings 
is heavily dumped by the surrounding plasma,
Eq.~(\ref{Y1}) is only valid for 
\be
\label{plasma}
G\mu>\left\{\begin{array}{lcl}
     \G\,m_\phi/(4\pi\,M_{\rm Pl}), & & \e> 1, \\
     \e\,\G\,m_\phi/(4\pi\,M_{\rm Pl}), & & \e< 1.\\
    \end{array} \right. \,
\ee
%

\section{Observational bounds and the detection of gravitational waves}

Having derived the abundance of dilatons (\ref{Y1}) we can now
apply different astrophysical constraints on the parameters
of cosmic strings. First, ultra-light dilatons are excluded by tests
of the gravitational inverse square law~\cite{Cavendish},
\be
\label{GISL}
m_\phi > 10^{-3} {\rm eV}\,.
\ee
Another limit comes from the bounds on the density of dark matter.
Dilatons interact with the gauge bosons through their mass terms in
the Lagrangian,  
$\mathcal{L}=(1/2)\alpha_F\,\kappa\phi\,F^2_{\mu\nu}$, where 
$\kappa=\sqrt{8\pi}/M_{\rm PL}$ and $\alpha_F$ parameterizes
deviations of the coupling at low-energies from the tree-level
coupling at the string scale. The exact value of $\alpha_F$ is
model-dependent, but generally of order one or
larger~\cite{Kaplan:2000hh}.  
Therefore dilatons decays into gauge bosons with the
lifetime~\cite{DamVil97} 
\be
\label{tau}
\tau_\phi\sim 3.3\times 10^{13}{\rm s}
\left(\frac{12}{N_F}\right)\left(\frac{\rm GeV}{m_\phi}\right)^3 \,, 
\ee
where $N_F$ is the number of gauge bosons with
masses below $m_\phi/2$ and we assumed $\alpha_F=1$. 

If the lifetime of dilatons is larger than the age of the universe, 
$\tau>t_0\sim 4\times 10^{17}$~s, the total density of dilatons is
bounded by the observed density of dark matter. According to the WMAP
observations, the relative fraction of dark matter is $\Omega_m
h^2=0.13$ \cite{Omega}, where $h=0.7$ parameterizes the Hubble constant.
Thus
\be
\label{OBound}
\Omega_\phi h^2=\frac{n_\phi\,m_\phi\,h^2}{\rho_{\rm cr}}<0.13,
\ee
where $\rho_{\rm cr}$ is the critical density of the universe.
Using (\ref{OBound}) one can easily obtain
\be
\label{YO}
Y_\phi <
\frac{0.13\,\rho_{\rm cr}}{h^2\,s(t_0)\,m_\phi}\simeq 4.5\times 10^{-10} 
\left(\frac{\rm GeV}{m_\phi}\right).
\ee
This constrain is valid for $\tau>t_0$ or dilaton masses
$m_\phi<0.1$~GeV. For lifetimes $\tau_\phi<t_0$ one can use
observations of the diffuse $\gamma$-ray background~\cite{gamma-ray}. 
Decaying dilatons produce photons with number
density $ n_\gamma\sim n_\phi(t_0)\, t_0/\tau_\phi$. Using this
relation and the approximate value for the upper bound of 
total energy density of photons with energy $>1$~MeV, 
$\rho_\gamma\sim 2\times 10^{-6}$ eV/cm$^{-3}$ \cite{gamma-ray},
we find
\be
\label{Yg1}
Y_\phi<\frac{\tau_\phi\,\rho_\gamma}{t_0\,s(t_0)\,m_\phi}\sim 
7\times 10^{-22}\,\left(\frac{\rm GeV}{m_\phi}\right)^4.
\ee

\begin{figure}[t]
\includegraphics[width=\textwidth]{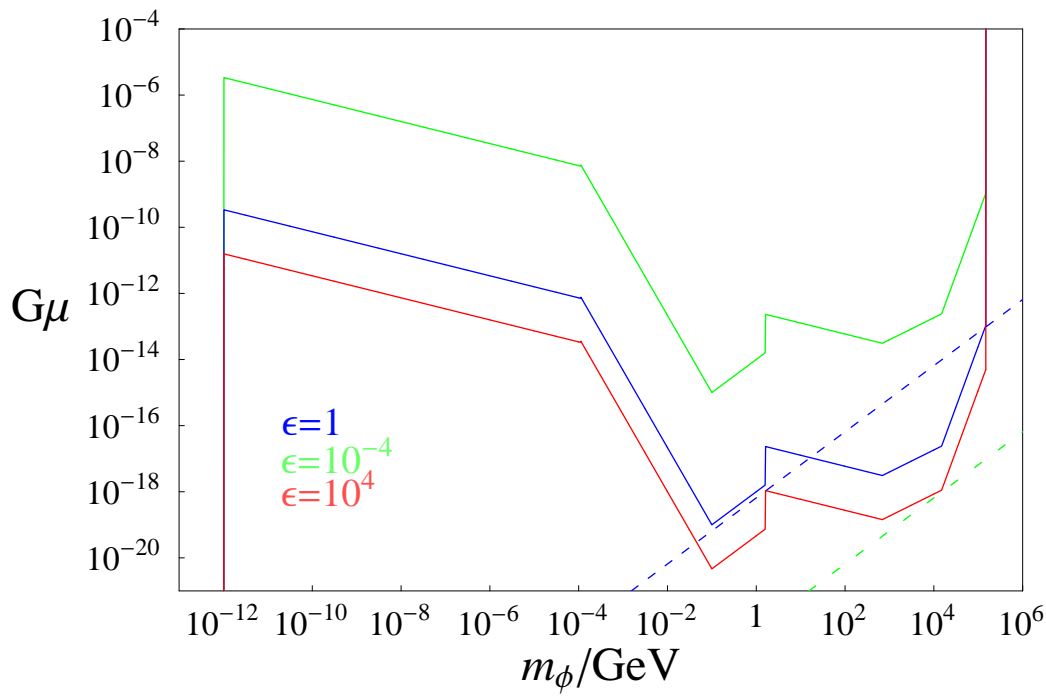}
\caption{\label{FigY} A log-log plot of upper bounds on $G\mu$ as 
function of $m_\phi$ for several values of $\e$; the other
parameters were chosen to have their ``standard'' values:
$\alpha=1$ and $p=1$. Below the dashed lines, plasma friction
invalidates our approximation.}
\end{figure}

For lifetimes $t_{\rm dec}<\tau<t_0$  of the dilaton (which corresponds
to dilaton masses $0.1\,{\rm GeV}<m_\phi<1.6$~GeV), where
$t_{\rm dec}\sim 10^{13}$ sec is the decoupling time, we
can also use the constraints from the density of $\gamma$-ray
background. The energy density of photons created due to 
decay of dilatons at a time $\tau$ is 
$\rho_\gamma(\tau)=m_\phi n_\phi(\tau)=m_\phi\,Y_\phi\,s(\tau)$.
Noting that the energy density of radiation scales with the time
as $\rho_\gamma\propto t^{8/3}$, we obtain
\be
\label{Yg2}
Y_\phi< 3\times 10^{-16}\left(\frac{m_\phi}{\rm GeV}\right).
\ee

For the lifetime of dilaton $\tau_\phi<t_{\rm dec}$ we obtain 
constraints on $Y_\phi$ from the abundances of $^4$He, $^3$He,
D and $^6$Li. Using the results of Ref.~\cite{BBH} we can
smoothly interpolate the results for decaying of relic particles
and approximate the maximal allowed abundance in the following way:
\be
  \label{YH}
    Y_\phi< \left\{\begin{array}{lcl}
     10^{-14} m_\phi^{-1}, & & 1.6\, {\rm GeV}<m_\phi<690\, {\rm GeV}, \\
     5.5\times 10^{-19}\,m_\phi^{1/2} & & 690\, {\rm GeV}<m_\phi<15\,
     {\rm TeV},\\
     9\times 10^{-38} m_\phi^5 & & 15\, {\rm TeV}\, <m_\phi<150\, {\rm
       TeV},\\
    \end{array} \right. \,
\ee
For very heavy dilaton masses, $m_\phi> 150$~TeV, there are no
limits on the abundance of dilatons.

\begin{figure}
\includegraphics[width=\textwidth]{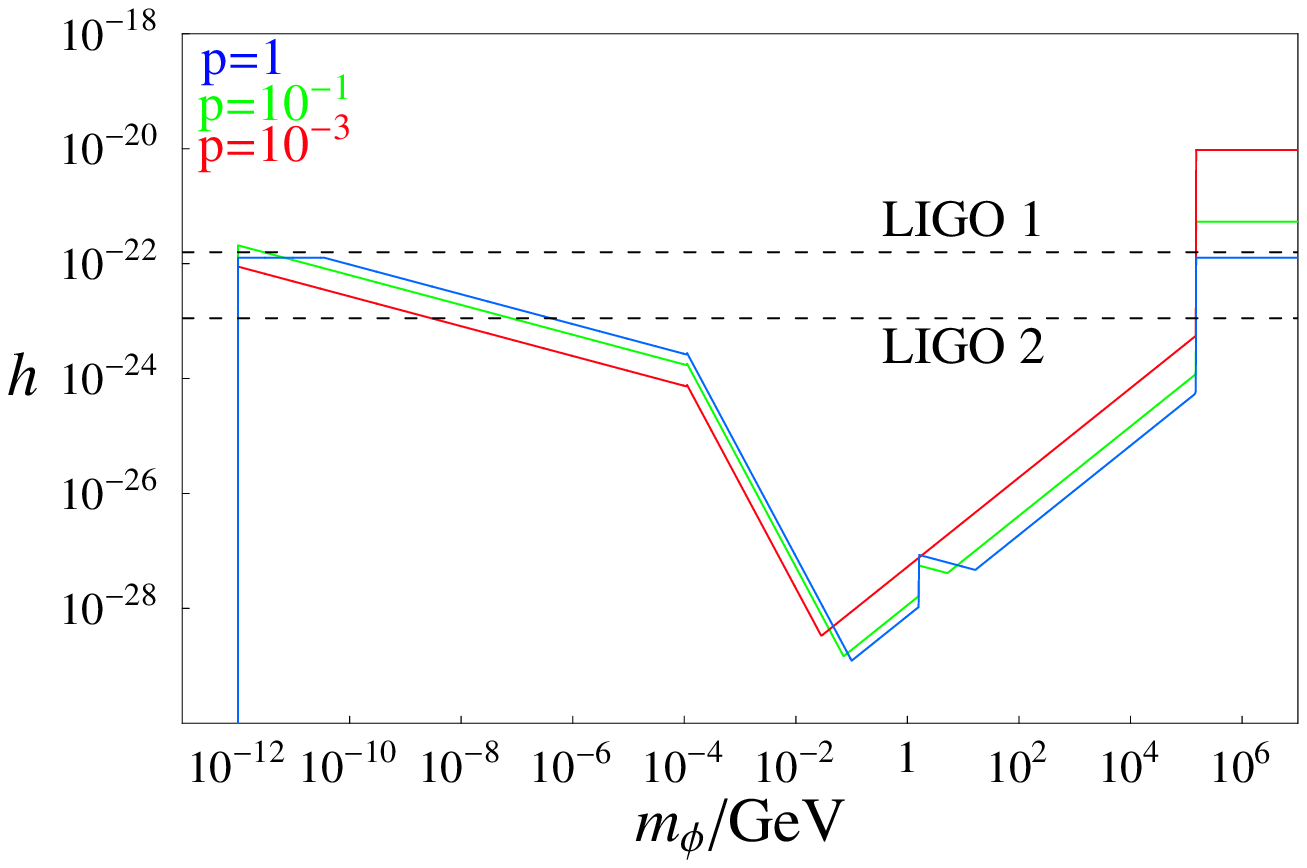}
\caption{\label{FigGWB_p} 
Maximal allowed values for the amplitude $h$ of the gravitational 
wave signal emitted in the LIGO frequency band, $f\sim 150$ Hz,
versus the dilaton mass $m_\phi$. The parameters of cosmic superstring
network are chosen to be $\e=1$ and $p=1$ (blue curve), 
$p=10^{-1}$ (green curve), $p=10^{-3}$ (red curve). The sensitivity
of LIGO is shown in the upper dashed line (initial configuration) and 
lower dashed line (advanced configuration).}
\end{figure}

\begin{figure}
\includegraphics[width=\textwidth]{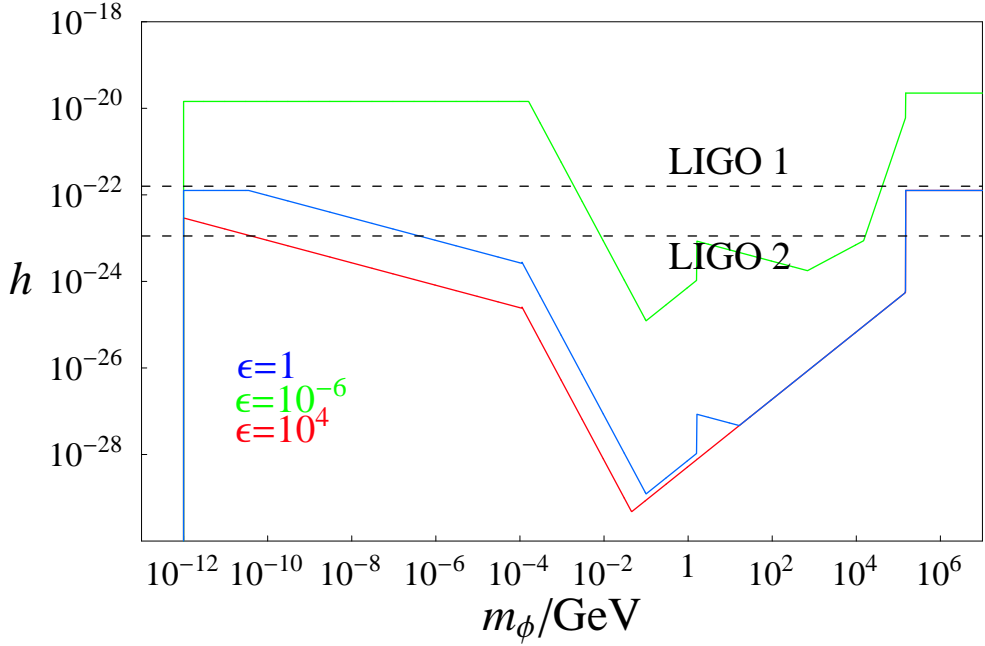}
\caption{\label{FigGWB_e} 
Maximal allowed values for the amplitude $h$ of the gravitational 
wave signal emitted in the LIGO frequency band, $f\sim 150$ Hz,
versus the dilaton mass $m_\phi$. The parameters of the cosmic
superstring network are chosen to be $p=1$ and $\e=1$ (blue curve), 
$\e=10^{-6}$ (green curve), $\e=10^4$ (red curve). The sensitivity
of LIGO is shown in the upper dashed line (initial configuration) and 
lower dashed line (advanced configuration).}
\end{figure}

Substituting the resulting expression for the dilaton abundance 
(\ref{Y1}) into the obtained constraints for $Y_\phi$ 
(\ref{YO}--\ref{YH}), one can find the 
constraints on the string tension $\mu$ as function 
of dilaton mass $m_\phi$, coupling constant $\alpha$, 
the probability  $p$ of reconnection of long strings, 
and the parameter $\e$, determining the size of the closed loops
produced by  the network.
For the ``standard'' values $p=1$ and $\e=1$ the resulting
Fig.~1 is similar to the one in Ref.~\cite{DamVil97}. 
The constraints on $G\mu$ are slightly more stringent, 
because of more precise observational data. However, the 
bounds on $G\mu$ may be strongly modified in comparison
with Ref.~\cite{DamVil97} by the multiplier 
$p^{-1}\,\e^{1/2}$ for $\e>1$ and $p^{-1}\,\e^{3/2}$ for $\e<1$.

Next we apply the obtained constraints on the values of $G\mu$ 
to examine the prospects to detect gravitational wave bursts (GWB)
from the cosmic superstring network. For each value of the dilaton 
mass $m_\phi$, we find the maximal allowed value $G\mu$ 
and then, using the results of \cite{DamVil04}, the maximal possible
signal for GWBs. We choose as frequency of the wave signal $150$~Hz,
i.e. the frequency range preferable for LIGO and assume one GWB event
per year. The results for the maximum possible 
amplitude of the GWB signals are shown in Fig.~\ref{FigGWB_p}
for different values of parameter $p$ and in
Fig.~\ref{FigGWB_e} for different values of $\e$. 
The sensitivity of the gravitational wave interferometer LIGO
(and its advanced configuration) is also shown in these figures.
We can see that the constraints from dilaton radiation significantly
restrict the prospects for the discovery of cosmic strings by the
detection of GWBs. In
particular, for the range of parameters $\e=1$ and $10^{-3}<p\le 1$,
GWBs from a string network would hardly be detected by LIGO for
dilaton masses $10^{-6}\ {\rm GeV}<m_\phi<10^5 GeV$. 
Only in the case that the typical size of closed string loops is much
smaller than in the 
standard scenario ($\e\ll 1$), the prospects to detect GWBs from
cosmic super strings improve: For instance, in the case of $p=1$ and
$\e=10^{-6}$ the detection of GRBs is possible for dilaton masses 
$10^{-12}{\rm GeV} \lsim m_\phi \lsim 10^{-2}$~GeV and for 
$m_\phi<10^4$~GeV. In the case of very small values of $\e$,
$\e<10^{-10}$, there are no limits on the GWBs amplitude coming from the 
dilaton abundance. We also checked the limits on
GWBs coming from the stochastic gravitational wave 
background~\cite{pulsar1}. However, this bounds is modified 
only by a factor $<3$ in the range of dilaton masses
$m_\phi<150$~TeV and cosmic string parameters,
$10^{-3}<p\le 1$ and $10^4<\e<10^{-6}$, therefore
the Figs.~\ref{FigGWB_p} and \ref{FigGWB_e}  change only 
slightly.

\section{Conclusions}

We have examined the emission of dilatons by a network of cosmic
superstrings. Depending on the particular mechanism of supersymmetry
breaking, one expects a dilaton mass between 10~eV and 10~TeV if
supersymmetry solves the hierarchy problem.
We have derived stringent limits on the string tension as function of
the dilation mass, cf.~Fig.~1, for the case of a non-standard
evolution of a string network.   
We have found that for dilaton mass in the favoured  range  between
10~eV and 10~TeV and values of the
string tension $10^{-12}\lsim G\mu\lsim 10^{-7}$ predicted in
Ref.~\cite{Jones:2003da}, the detection of a gravitational wave signal
from cosmic superstrings is only possible when the evolution of the
string network deviates strongly from the standard case.

\section*{Acknowledgments}
We are grateful to Alex Vilenkin for useful comments.
This work was supported by the Deut\-sche
For\-schungs\-ge\-mein\-schaft within the Emmy Noether program and
the Russian Foundation for Basic Research, grant 04-02-16757-a.


\end{document}